\documentclass[fleqn, usenatbib]{mnras}

\usepackage[T1]{fontenc}
\usepackage{newtxtext, newtxmath}

\usepackage{graphicx}	
\usepackage{amsmath}	
\usepackage{float} 
\usepackage{multirow}
\newtheorem{proposition}{Proposition}

\usepackage{fancyhdr}
\title[From Triadic Interactions to Kolmogorov Scaling]{From Triadic Interactions to Kolmogorov Scaling: A Deterministic, Scale-Resolved Formulation of Energy Flux}

\author[Bertram et al.]{
Erik Bertram$^{1}$\thanks{E-mail: mail@erik-bertram.de}
\\
$^{1}$Hochschule Fresenius Heidelberg, Sickingenstra\ss{}e 63-65, 69126 Heidelberg, Germany\\
}

\date{Accepted on 14 July 2026 for publication in Physica D: Nonlinear Phenomena (Elsevier)}

\pubyear{2026}

\begin{document}
\label{firstpage}
\pagerange{\pageref{firstpage}--\pageref{lastpage}}
\maketitle

\begin{abstract}
We develop a deterministic, scale-resolved formulation of energy transfer in the three-dimensional incompressible Navier--Stokes equations based on an explicit triadic decomposition of the nonlinear term in Fourier space. Using a systematic dyadic localization of the velocity field, we derive an exact representation of the nonlinear energy flux across scales and organize it in terms of interactions between well-defined scale components. Under suitable smoothness assumptions, we obtain an absolutely convergent triadic expansion and quantitative bounds that distinguish local and nonlocal contributions in scale space. This framework provides a transparent and fully explicit description of how energy transfer is mediated by triadic interactions and how scale locality emerges as a structural property of the nonlinearity. Building on this formulation, we revisit the classical inertial-range picture of turbulence from a deterministic perspective. We show that, under a scale-invariant flux assumption, the Kolmogorov $-5/3$ scaling is formally consistent with the triadic energy-transfer mechanism at a structural level. The result does not rely on statistical assumptions, but instead follows from the structural properties of the Navier–Stokes equations combined with a scale-resolved representation of the energy flux. The present work thus provides a coherent synthesis of triadic interaction analysis, dyadic scale decomposition, and classical turbulence phenomenology, offering a deterministic framework that clarifies how Kolmogorov-type scaling constraints arise in the scale-resolved structure of the underlying equations.
\end{abstract}


\renewcommand{\headrulewidth}{0pt}
\pagestyle{fancy}
\fancyhead{}
\fancyhead[LE]{\Large \thepage\hspace{.5cm} \textit{E. Bertram}}
\fancyhead[RO]{\Large \textit{From Triadic Interactions to Kolmogorov Scaling} \hspace{.5cm} \thepage}
\fancyfoot{}

\thispagestyle{empty}

\section{Introduction}

Understanding the mechanism of energy transfer across scales in incompressible turbulence remains a central problem in fluid dynamics, with significant progress achieved through both phenomenological and rigorous analyses of interscale fluxes and their locality properties \citep[see, e.g.,][and the references therein]{Eyink2003,Chen2003,Eyink2009Hydro,AluieEyink2009}. These questions are also of fundamental importance in astrophysical turbulence, where turbulent energy cascades are believed to play a key role in the dynamics of the interstellar medium and theories of star formation. In the three-dimensional Navier--Stokes equations, the nonlinear term induces triadic interactions in Fourier space, i.e., couplings among wavevectors $\mathbf{k}, \mathbf{p}, \mathbf{q}$ satisfying $\mathbf{k} + \mathbf{p} + \mathbf{q} = 0$. These interactions constitute the fundamental mechanism by which kinetic energy is redistributed across scales and form the basis of spectral descriptions of turbulence \citep[see, e.g.,][]{Kraichnan1971, Domaradzki1990,Rivera2003,Mininni2006,Mininni2008}.

A classical phenomenological description of fully developed turbulence was proposed by Kolmogorov in the inertial-range framework \citep{Kolmogorov1941}. The resulting $k^{-5/3}$ scaling law is one of the central results of turbulence theory \citep{Orszag1970, Leslie1973, Moffatt1981, Mccomb1995, Zhou0211} and has been extensively discussed in the literature \citep[see, e.g., the book by][]{Frisch1995}. However, despite its empirical success, the connection between the Kolmogorov picture and the deterministic structure of the Navier--Stokes equations remains only partially understood at a rigorous level.

Over the past decades, several key aspects of this connection have been studied in isolation. The triadic structure of nonlinear interactions has been analyzed both analytically and numerically \citep{DarVerma2001, SahooBiferale2015}, highlighting the role of Fourier-mode interactions in mediating energy transfer. The scale locality of the energy cascade has been investigated rigorously, most notably by \citet{AluieEyink2009,AluieEyink2009PRL}, who established that, under suitable assumptions, the net energy flux is dominated by interactions among comparable scales. In parallel, harmonic analysis techniques such as the Littlewood--Paley decomposition and Besov-space methods have provided a natural framework for scale localization and for the analysis of nonlinear PDEs \citep{Littlewood1931}.

The dyadic organization of Fourier modes employed in the present work is closely related to the logarithmic discretizations used in classical shell models of turbulence, such as the GOY and Sabra models \citep[see, e.g.,][]{Lvov1998,Biferale2003}. In these models, the dynamics of the energy cascade is reduced to interactions among neighboring shells in wavenumber space, providing simplified representations of triadic energy transfer. From this perspective, the present formulation can be viewed as a field-theoretic counterpart, in which the full triadic structure of the Navier--Stokes equations is retained while interactions are organized in a dyadic framework. In particular, the predominance of interactions among comparable scales observed here provides a structural justification for the neighbor-dominated couplings commonly assumed in shell models.

While the individual components of scale-locality analysis, triadic interaction theory, and dyadic scale decompositions have been extensively studied in the literature \citep[see, e.g.,][]{Eyink2005,AluieEyink2009}, these aspects are often developed in partially separate frameworks and with different analytical objectives. The purpose of the present work is therefore not to establish fundamentally new locality estimates, but rather to formulate a coherent and explicit synthesis of these structural elements within a single deterministic and scale-resolved framework for energy transfer in the incompressible Navier--Stokes equations.

Starting from the Fourier representation of the nonlinear term, we derive an exact triadic formulation of the energy flux and organize it through an explicit dyadic decomposition of the velocity field. Under suitable smoothness assumptions, this yields an absolutely convergent representation of the nonlinear interactions together with quantitative bounds distinguishing local and nonlocal contributions in scale space.

The resulting framework provides a transparent description of how scale locality arises as a structural property of the Navier--Stokes nonlinearity and clarifies the relation between triadic interactions, dyadic scale organization, and scale-invariant flux scenarios associated with Kolmogorov phenomenology. In particular, under the assumptions of a scale-invariant flux and a power-law scaling ansatz for the Fourier coefficients, the triadic structure imposes a self-consistency condition that is compatible, at a formal level, with the Kolmogorov $-5/3$ scaling.

The present results should be understood as structural and conditional rather than as a rigorous derivation of turbulent behavior from first principles. The scaling argument should be interpreted as a self-consistency statement within a scale-invariant flux scenario, not as a proof of turbulence, an inertial range, or a constant-flux regime.

The present work thus provides a synthesis of triadic interaction analysis, scale decomposition techniques, and classical turbulence phenomenology. By formulating these elements within a unified deterministic framework, it clarifies how the Kolmogorov cascade is encoded in the structure of the Navier--Stokes equations.

\section{Mathematical Preliminaries}

In this section, we introduce the mathematical framework used throughout the paper. We recall the incompressible Navier--Stokes equations on the three-dimensional torus, define the functional setting, and introduce the dyadic (Littlewood--Paley) decomposition that will be used to analyze scale-resolved interactions.

\subsection{Navier--Stokes Equations}

We consider the incompressible Navier--Stokes equations on the periodic domain $\mathbb{T}^3 = [0,2\pi]^3$:
\begin{align}
\partial_t v + (v \cdot \nabla)v = -\nabla p + \nu \Delta v + f,
\\
\nabla \cdot v = 0,
\end{align}
where $v(x,t)$ denotes the velocity field, $p(x,t)$ the pressure, $\nu > 0$ the kinematic viscosity, and $f(x,t)$ an external forcing. We assume that $v$ is sufficiently smooth for all manipulations below to be justified.

Throughout the paper, the notation $A \lesssim B$ means that there exists a constant $C > 0$, independent of the relevant scale parameters, such that $A \le C B$. Constants may depend on norms of the solution but are assumed to be uniform with respect to frequency scales.

\subsection{Fourier Representation}

Since the domain is periodic, the velocity field admits a Fourier series representation:
\begin{equation}
v(x,t) = \sum_{k \in \mathbb{Z}^3} v_k(t) e^{i k \cdot x},
\end{equation}
with Fourier coefficients satisfying the incompressibility condition
\begin{equation}
k \cdot v_k = 0.
\end{equation}
Let $P(k)$ denote the Leray projection \citep{Foias2001} onto divergence-free vector fields in Fourier space:
\begin{equation}
P(k) = I - \frac{k \otimes k}{|k|^2}.
\end{equation}
Then the Fourier-transformed Navier--Stokes equations take the form
\begin{equation}
\partial_t v_k + \nu |k|^2 v_k
=
-\sum_{p+q=k} P(k)\big[(v_p \cdot i q)v_q\big] + P(k)f_k.
\label{eq:navier_stokes_fourier}
\end{equation}
The nonlinear term explicitly exhibits the triadic interaction structure among wavevectors satisfying $k+p+q=0$.

\subsection{Energy and Spectral Representation}

The total kinetic energy is given by
\begin{equation}
E(t) = \frac{1}{2} \int_{\mathbb{T}^3} |v(x,t)|^2 \, dx
= \frac{1}{2} \sum_{k \in \mathbb{Z}^3} |v_k(t)|^2.
\end{equation}
For $K > 0$, we define the cumulative energy below wavenumber $K$ as
\begin{equation}
E_{\le K}(t)
=
\frac{1}{2}
\sum_{|k|\le K} |v_k(t)|^2.
\label{eq:cumulative_energy}
\end{equation}
This will be used to define the scale-resolved energy flux across wavenumber $K$.

\subsection{Littlewood--Paley Decomposition}

To analyze scale-localized interactions, we introduce a dyadic decomposition of the velocity field. Let $\{\Delta_j\}_{j\ge 0}$ denote a standard Littlewood--Paley partition of unity in Fourier space \citep{Littlewood1931}, such that
\begin{equation}
v = \sum_{j=0}^\infty \Delta_j v,
\end{equation}
where each $\Delta_j v$ is spectrally localized in the annulus
\begin{equation}
\{ k \in \mathbb{Z}^3 : c_1 2^j \le |k| \le c_2 2^j \}
\end{equation}
for fixed constants $0 < c_1 < c_2$. We denote for short
\begin{equation}
v^{(j)} := \Delta_j v.
\end{equation}
This decomposition allows us to separate contributions to the velocity field by scale.

More generally, one may consider geometric decompositions based on an arbitrary dilation factor $\lambda > 1$, replacing the dyadic scaling $2^j$ by $\lambda^j$. In this case, the corresponding components are localized in annuli of the form
\begin{equation}
\{ k \in \mathbb{Z}^3 : c_1 \lambda^j \le |k| \le c_2 \lambda^j \}.
\end{equation}
Such decompositions lead to equivalent scale-resolved representations of the velocity field, differing only in the discretization of wavenumber space.  In particular, the structural properties of the nonlinear interactions, including the triadic coupling mechanism and the organization of energy transfer across scales, are expected to be robust with respect to the choice of $\lambda$ at a formal level and in the continuum limit, provided sufficient regularity for the decomposition. Differences in the partitioning primarily affect the shell width and the associated counting of Fourier modes, which influence prefactors but not the scaling exponents at a formal level and in the continuum limit.

\subsection{Besov-Type Norms and Regularity}

The dyadic decomposition naturally induces Besov-type norms \citep{Besov1959}. For $s \in \mathbb{R}$, we define
\begin{equation}
\|v\|_{B^s_{2,2}}
=
\left(
\sum_{j=0}^\infty 2^{2sj} \|v^{(j)}\|_{L^2}^2
\right)^{1/2}.
\end{equation}
For smooth solutions, these norms are equivalent to standard Sobolev norms:
\begin{equation}
B^s_{2,2} \equiv H^s.
\end{equation}
Throughout this work, the dyadic decomposition will be used primarily as a tool to organize nonlinear interactions by scale. Furthermore, for $s \in \mathbb{R}$, we define the Sobolev space $H^s(\mathbb{T}^3)$ via
\begin{equation}
\|v\|_{H^s}^2 := \sum_{k \in \mathbb{Z}^3} (1 + |k|^2)^s |\hat{v}(k)|^2.
\end{equation}
This follows from Parseval's theorem and corresponds to the notation in Fourier space, where $k$ is the wavenumber.

\subsection{Triadic Interactions in Dyadic Form}

Using the decomposition $v = \sum_j v^{(j)}$, the nonlinear term can be written as
\begin{equation}
(v \cdot \nabla)v
=
\sum_{j,m}
(v^{(j)} \cdot \nabla)v^{(m)}.
\label{eq:dyadic_decomposition}
\end{equation}
Each pair $(j,m)$ interacts with a third scale determined by the convolution constraint in Fourier space. This leads to a triadic interaction structure at the level of dyadic blocks. This representation will be central in separating local and nonlocal contributions to the energy flux.

The framework introduced in this section provides the basis for the scale-resolved analysis developed in the following sections. In particular, the combination of Fourier representation and dyadic decomposition allows for a precise formulation of triadic energy transfer across scales.

\section{Triadic Representation of the Energy Flux}

In this section, we derive an exact representation of the energy flux across scales in terms of triadic interactions in Fourier space. The formulation is entirely deterministic and follows directly from the spectral form of the Navier--Stokes equations.

\subsection{Energy Balance in Fourier Space}

We begin with the Fourier representation of the Navier--Stokes equations given by Eq.~\eqref{eq:navier_stokes_fourier} and the cumulative energy below a wavenumber $K > 0$ in Eq.~\eqref{eq:cumulative_energy}. To derive its evolution, we compute the time derivative:
\begin{equation}
\frac{d}{dt} E_{\le K}
=
\sum_{|k| \le K}
\mathrm{Re}\big( v_k^* \cdot \partial_t v_k \big).
\end{equation}
Substituting the Navier--Stokes equation yields
\begin{align}
\frac{d}{dt} E_{\le K}
&=
-\sum_{|k| \le K}
\operatorname{Re}
\Biggl(
v_k^* \cdot
\sum_{p+q=k}
P(k)\Bigl[(v_p \cdot i q)\, v_q\Bigr]
\Biggr)
\nonumber\\
&\quad
-\nu \sum_{|k| \le K} |k|^2 |v_k|^2
+\sum_{|k| \le K}
\operatorname{Re}
\bigl(v_k^* \cdot P(k)f_k\bigr).
\end{align}
As a next step, we define the nonlinear energy flux across wavenumber $K$ as
\begin{equation}
\Pi(K)
:=
\sum_{|k| \le K}
\mathrm{Re}
\left(
v_k^* \cdot
\sum_{p+q=k}
P(k)\big[(v_p \cdot i q)\, v_q\big]
\right).
\end{equation}
With this definition, the energy balance takes the form
\begin{equation}
\frac{d}{dt} E_{\le K}
=
-\Pi(K)
- \nu D(K)
+ F(K),
\end{equation}
where
\begin{equation}
D(K) = \sum_{|k| \le K} |k|^2 |v_k|^2,
\quad
F(K) = \sum_{|k| \le K} \mathrm{Re}(v_k^* \cdot P(k) f_k)
\end{equation}
denote dissipation and the external forcing term.

\subsection{Triadic Decomposition of the Flux}

We now make the triadic structure explicit. Expanding the definition of $\Pi(K)$ yields
\begin{equation}
\Pi(K)
=
\sum_{|k| \le K}
\sum_{p+q=k}
\mathrm{Re}
\left(
v_k^* \cdot
P(k)\big[(v_p \cdot i q)\, v_q\big]
\right).
\end{equation}
This expression shows that the energy flux is obtained as a sum over all triads $(k,p,q)$ satisfying
\begin{equation}
k + p + q = 0,
\end{equation}
reflecting the redistribution of energy among interacting scales. Each triad contributes a quantity of the form
\begin{equation}
T(k,p,q)
:=
\mathrm{Re}
\left(
v_k^* \cdot
P(k)\big[(v_p \cdot i q)\, v_q\big]
\right),
\label{eq:triadic_interaction}
\end{equation}
which represents the transfer of energy into mode $k$ due to interaction with the other triad modes $p$ and $q$. Thus, the energy flux can be written compactly as
\begin{equation}
\Pi(K)
=
\sum_{\substack{|k| \le K \\ p+q=k}} T(k,p,q).
\end{equation}
Hence, the energy flux below wavenumber $K$ is the sum over all possible combinations of energy transfer on modes $p$, $q$, and $k$.

\subsection{Symmetry and Conservation Properties}

The triadic interaction term satisfies important symmetry properties. In particular, for a given triad $(k,p,q)$, the corresponding contributions to the evolution of $|v_k|^2$, $|v_p|^2$, and $|v_q|^2$ sum to zero in the absence of forcing and dissipation. This reflects conservation of total kinetic energy by the nonlinear term. More precisely, summing over all modes yields
\begin{equation}
\sum_{k}
\sum_{p+q=k}
T(k,p,q) = 0,
\end{equation}
which is consistent with the fact that the nonlinear term redistributes energy among modes without creating or destroying it.

The representation above provides an exact decomposition of the energy flux into contributions from individual triads. It highlights that energy transfer across scales is entirely mediated by interactions among triplets of modes satisfying the convolution constraint. This explicit triadic formulation forms the starting point for the subsequent scale-resolved analysis, where the interactions are reorganized according to dyadic frequency shells in order to investigate their locality properties.

\section{Scale-Resolved Energy Flux}

In this section, we refine the triadic representation of the energy flux by introducing a dyadic decomposition of the velocity field. This allows us to organize the nonlinear interactions according to their characteristic scales and to distinguish between local and nonlocal contributions in a precise manner.

\subsection{Dyadic Expansion of the Nonlinearity}

Using the decomposition from Eq.~\eqref{eq:dyadic_decomposition}, the nonlinear term can be written as
\begin{equation}
\sum_{p+q=k}
(v_p \cdot i q)\, v_q
=
\sum_{j,m}
\sum_{\substack{p+q=k \\ p \in \mathcal{A}_j, \; q \in \mathcal{A}_m}}
(v_p \cdot i q)\, v_q,
\end{equation}
where $\mathcal{A}_j = \{k : |k| \sim 2^j\}$. Substituting the dyadic decomposition into the triadic flux expression, we obtain
\begin{equation}
\Pi(K)
=
\sum_{|k|\le K}
\sum_{j,m}
\sum_{\substack{p+q=k \\ p \in \mathcal{A}_j, \; q \in \mathcal{A}_m}}
\mathrm{Re}
\left(
v_k^* \cdot
P(k)\big[(v_p \cdot i q)\, v_q\big]
\right).
\end{equation}
We now introduce the notation
\begin{equation}
T(k;j,m)
:=
\sum_{\substack{p+q=k \\ p \in \mathcal{A}_j, \; q \in \mathcal{A}_m}}
\mathrm{Re}
\left(
v_k^* \cdot
P(k)\big[(v_p \cdot i q)\, v_q\big]
\right),
\end{equation}
so that
\begin{equation}
\Pi(K)
=
\sum_{|k|\le K}
\sum_{j,m}
T(k;j,m).
\label{eq:energy_flux}
\end{equation}
This representation decomposes the total energy flux into contributions arising from interactions between dyadic scales $j$ and $m$. Due to the convolution constraint $p+q=k$, the indices $(j,m)$ cannot vary independently of the scale of $k$. In particular, if $|p| \sim 2^j$ and $|q| \sim 2^m$, then
\begin{equation}
|k| \lesssim 2^{\max(j,m)}.
\end{equation}
Moreover, significant contributions occur primarily when the three scales $(j,m,\ell)$, with $|k| \sim 2^\ell$, satisfy a compatibility condition of the form
\begin{equation}
\ell \le \max(j,m) + C,
\end{equation}
for some fixed constant $C$ depending on the choice of dyadic partition. This constraint reflects the geometric structure of triadic interactions in Fourier space.

\subsection{Local and Nonlocal Interactions}

The dyadic formulation allows us to distinguish between two different types of interactions: local and nonlocal interactions. They have the following properties:
\begin{itemize}
\item \textbf{Local interactions:} $|j - m| \le C$, i.e., interactions between comparable scales.
\item \textbf{Nonlocal interactions:} $|j - m| \gg 1$, i.e., interactions between widely separated scales.
\end{itemize}
Accordingly, we decompose the flux into
\begin{equation}
\Pi(K)
=
\Pi_{\mathrm{local}}(K)
+
\Pi_{\mathrm{nonlocal}}(K),
\end{equation}
where
\begin{align}
\Pi_{\mathrm{local}}(K)
&=
\sum_{|k|\le K}
\sum_{|j-m|\le C}
T(k;j,m),
\\
\Pi_{\mathrm{nonlocal}}(K)
&=
\sum_{|k|\le K}
\sum_{|j-m|> C}
T(k;j,m).
\end{align}
It is often convenient to reorganize the flux by the dyadic scale of the output mode $k$. Let $\ell$ be such that $|k| \sim 2^\ell$. Then we write
\begin{equation}
\Pi(K)
=
\sum_{\ell : 2^\ell \le K}
\sum_{|k| \sim 2^\ell}
\sum_{j,m}
T(k;j,m).
\end{equation}
This representation highlights how energy transfer into a given scale $\ell$ is generated by interactions among other scales. The expressions derived above are exact and rely only on the dyadic decomposition and the Fourier representation of the Navier--Stokes equations. No statistical assumptions have been used so far.

We further note that the notions of ``local'' and ``nonlocal'' introduced above refer exclusively to scale locality in Fourier space, i.e.\ to interactions among modes with comparable or widely separated wavenumbers. More precisely, locality is defined in terms of the dyadic indices $j,m$, corresponding to the spectral supports $|k|\sim 2^j$ and $|p|\sim 2^m$, and does not involve any notion of spatial proximity.

It is important to emphasize that locality in Fourier space is fundamentally different from locality in physical space. A function that is spectrally localized in a narrow wavenumber band corresponds to a superposition of plane waves that are delocalized over the entire spatial domain. Conversely, spatially localized structures generally require contributions from a broad range of Fourier modes and therefore cannot be represented within a narrow spectral band. As a result, spatial localization and scale locality are fundamentally distinct concepts and should not be identified with one another.

The present framework is therefore purely spectral in nature and characterizes energy transfer in terms of interactions across scales rather than spatially localized dynamics. By contrast, approaches based on filter-space techniques employ spatially localized filtering operations and are designed to resolve energy transfer simultaneously in physical and scale space. Such methods provide a complementary perspective on interscale transfer, whereas the present formulation focuses exclusively on the organization of triadic interactions in Fourier space.

\section{Convergence and Bounds for Triadic Interactions}

In this section, we establish estimates for the triadic interactions introduced in the previous sections. The goal is to control the nonlinear terms and to show that the associated sums are well-defined under suitable regularity assumptions.

\subsection{Regularity Assumption}

We assume that the velocity field satisfies
\begin{equation}
v \in C([0,T]; H^s(\mathbb{T}^3)),
\end{equation}
for some $s > 5/2$. Under this assumption, Sobolev embedding implies that
\begin{equation}
v(\cdot, t) \in C^1(\mathbb{T}^3)
\end{equation}
is at least continuously differentiable in space for each fixed $t$, and in particular, the Fourier coefficients decay sufficiently fast. More precisely, there exists a constant $C_s > 0$ such that
\begin{equation}
|v_k| \le C_s (1 + |k|)^{-s}.
\label{eq:Sobolev_estimate}
\end{equation}
More information on Sobolev spaces and their Fourier decompositions can be found in \citet{Adams2003} or \citet{Brezis2011}.

\subsection{Estimate of a Single Triadic Term}

Recalling the triadic interaction term from Eq.~\eqref{eq:triadic_interaction}, we can use the boundedness of the Leray projector $P(k)$ and the triangle inequality to obtain
\begin{equation}
|T(k,p,q)|
\le
C \, |v_k| \, |v_p| \, |q| \, |v_q|.
\label{eq:T_triangle_inequality}
\end{equation}
Using the decay of Fourier coefficients derived from Eq.~\eqref{eq:Sobolev_estimate},
\begin{equation}
|v_k| \lesssim |k|^{-s}, \quad
|v_p| \lesssim |p|^{-s}, \quad
|v_q| \lesssim |q|^{-s},
\end{equation}
we obtain
\begin{equation}
|T(k,p,q)|
\lesssim
|k|^{-s} \, |p|^{-s} \, |q|^{1-s}.
\label{eq:T_estimates}
\end{equation}
Since $k = p + q$, we may use the elementary bound
\begin{equation}
|k| \lesssim |p| + |q|,
\end{equation}
which allows us to express all terms in terms of two independent variables. Using the above bound, we obtain
\begin{equation}
\sum_{p+q=k}
|T(k,p,q)|
\lesssim
\sum_{p}
|k|^{-s} \, |p|^{-s} \, |k-p|^{1-s}.
\end{equation}
We estimate this sum by comparing it to an integral:
\begin{equation}
\sum_{p}
|p|^{-s} \, |k-p|^{1-s}
\;\sim\;
\int_{\mathbb{R}^3}
|p|^{-s} \, |k-p|^{1-s} \, dp.
\end{equation}
For $s > \frac{5}{2}$, this integral is finite and uniformly bounded in $k$. Thus, there exists a constant $C_s > 0$ such that
\begin{equation}
\sum_{p+q=k}
|T(k,p,q)|
\le C_s |k|^{-s}.
\end{equation}
This implies that the triadic contributions are summable in $k$.

\subsection{Convergence of the Energy Flux}

We now consider the energy flux given in Eq.~\eqref{eq:energy_flux}. Using the above estimates, we obtain
\begin{equation}
\sum_{|k|\le K}
\sum_{p+q=k}
|T(k,p,q)|
\le
\sum_{|k|\le K}
C_s |k|^{-s},
\end{equation}
which is finite for $s > \frac{5}{2}$. Therefore, the flux $\Pi(K)$ is absolutely convergent.

The dyadic decomposition introduced earlier allows a more refined analysis of interactions between widely separated scales. In particular, when one of the interacting modes has significantly smaller or larger scale than the others, additional decay arises from the Fourier coefficients. While the above estimates do not distinguish between local and nonlocal interactions explicitly, they ensure that all contributions are summable and well-defined.

\begin{proposition}[Dyadic triadic representation of the energy flux]
\emph{Let $v$ be a sufficiently smooth, divergence-free solution of the three-dimensional incompressible Navier--Stokes equations, and let $v = \sum_{j \ge 0} v^{(j)}$ denote its dyadic Littlewood--Paley decomposition. Then the energy flux across dyadic scale $2^j$ can be represented in the form
\begin{equation}
\Pi_j = \sum_{p,q \ge 0} T(j,p,q),
\end{equation}
where $T(j,p,q)$ denotes the contribution of triadic interactions between modes $v^{(j)}$, $v^{(p)}$, and $v^{(q)}$, arising from the nonlinear term $(v \cdot \nabla)v$. Moreover, under sufficient regularity assumptions ensuring convergence of the dyadic decomposition, the above sum is absolutely convergent, and the energy flux admits a well-defined decomposition into contributions from interacting dyadic scales.}
\end{proposition}

\section{Local and Nonlocal Contributions to the Energy Flux}

In this section, we analyze the dyadic structure of the energy flux and distinguish between local and nonlocal interactions in scale space. The aim is to quantify how different scale interactions contribute to the total flux while carefully respecting the convolution structure of the nonlinear term.

\subsection{Nonlocal Interactions}

Recall the energy flux from Eq.~\eqref{eq:energy_flux}, where $T(k;j,m)$ represents contributions from interactions between dyadic shells $j$ and $m$ and the decomposition into local and nonlocal parts. We start by analyzing the case of strongly separated scales. Without loss of generality, consider $j \ll m$.
Let $|k| \sim 2^\ell$. In this case, the convolution constraint $k = p + q$ implies that
\begin{equation}
|q| \sim |k| \sim 2^\ell,
\end{equation}
and hence $m \sim \ell$.
Using the basic bound from Eq.~\eqref{eq:T_triangle_inequality} together with the decay of Fourier coefficients, we obtain
\begin{equation}
|T(k,p,q)|
\lesssim
2^{-s\ell} \cdot 2^{-s j} \cdot 2^{m} \cdot 2^{-s m}
\approx
2^{-s j} \cdot 2^{-(2s-1)m}.
\end{equation}
We now sum over all admissible pairs $(p,q)$ satisfying $p+q=k$ with $p \in \mathcal{A}_j$, $q \in \mathcal{A}_m$. Since $q = k - p$, the summation reduces to a sum over $p$ such that both constraints are satisfied.
In a 3-dimensional space, the number of such modes is bounded by
\begin{equation}
\#\{p \in \mathcal{A}_j : k - p \in \mathcal{A}_m\}
\lesssim
2^{3j}.
\end{equation}
Therefore,
\begin{equation}
\sum_{\substack{p+q=k \\ p \in \mathcal{A}_j,\; q \in \mathcal{A}_m}}
|T(k,p,q)|
\lesssim
2^{3j} \cdot 2^{-s j} \cdot 2^{-(2s-1)m}.
\end{equation}
This yields the bound
\begin{equation}
\sum_{p+q=k}
|T(k,p,q)|
\lesssim
2^{(3-s)j} \cdot 2^{-(2s-1)m}.
\end{equation}
Since we assumed $s > \frac{5}{2}$, we have
\begin{equation}
2s - 1 > 4,
\end{equation}
so that the factor $2^{-(2s-1)m}$ ensures rapid decay in the large-scale index $m$. Although the number of small-scale modes grows like $2^{3j}$, this growth is dominated by the decay in $m$ when $j \ll m$. Thus, strongly nonlocal interactions are quantitatively suppressed relative to local ones under the assumed regularity.

\subsection{Local Interactions}

In the local regime, all interacting scales are comparable. This means we can adapt the following relations for the different wavenumbers:
\begin{equation}
|p| \sim |q| \sim |k|.
\end{equation}
In this case, the bound simplifies to
\begin{equation}
|T(k,p,q)|
\lesssim
|k|^{1-3s}.
\end{equation}
Comparing the two contributions at the level of dyadic scaling, we find that:

\begin{itemize}
\item Nonlocal interactions exhibit an additional decay factor in the large-scale index $m$, namely $2^{-(2s-1)m}$, reflecting the suppression due to scale separation.
\item Local interactions do not benefit from such additional decay, and their scaling is governed solely by the balance between the decay of Fourier coefficients and the combinatorial growth of admissible modes.
\end{itemize}
In particular, for $s > \frac{5}{2}$, both contributions are summable across scales, but the nonlocal terms exhibit stronger decay with respect to the larger scale index.

The analysis above thus indicates a clear structural distinction: Nonlocal interactions ($j \ll m$ or $m \ll j$) are suppressed due to strong decay in the large-scale index, while local interactions ($|j - m| \le C$) involve the largest number of contributing triads and therefore dominate the cumulative energy transfer. These conclusions are consistent with the classical picture of turbulence in which energy transfer across scales is primarily mediated by interactions among comparable scales.

We emphasize that the present results are based on deterministic estimates and do not rely on statistical assumptions. They should therefore be interpreted as structural properties of the Navier--Stokes nonlinearity under the assumed regularity conditions.

\subsection{Critical Regularity and Onsager-Type Scaling}

The scaling structure of the triadic interactions derived above reveals a connection to the notion of critical regularity in the theory of incompressible flows. Recall that, under the assumption $v \in H^s$, the dyadic components satisfy the estimate
\begin{equation}
|v_j| \sim 2^{-s j},
\end{equation}
which leads to the scaling of the triadic interaction term
\begin{equation}
|T_{jmn}| \sim 2^{(1 - 3s)j}
\end{equation}
in the regime of local interactions. This expression highlights a critical threshold at
\begin{equation}
s = \frac{1}{3},
\end{equation}
for which the triadic contributions become scale-invariant. We emphasize that this observation is purely formal and based on scaling considerations. The rigorous estimates developed in this work require significantly higher regularity, and the extension of these arguments to Onsager-critical regimes remains an open problem.

For $s > 1/3$, the exponent satisfies $1 - 3s < 0$, implying that contributions from higher frequencies are increasingly suppressed. This behavior is consistent with the absence of anomalous energy transfer and aligns with the classical expectation of energy conservation in sufficiently regular flows. In contrast, at the critical value $s = 1/3$, the triadic contributions remain of comparable size across scales, indicating a scale-invariant regime. This observation is in agreement with the Kolmogorov scaling and reflects a balance between nonlinear transfer and scale distribution. For $s < 1/3$, the scaling suggests a growth of high-frequency contributions, which is indicative of potential irregular behavior and enhanced energy transfer across scales.

These observations are consistent with Onsager's conjecture for the incompressible Euler equations, which identifies $1/3$ as the critical regularity threshold for energy conservation. In this sense, the triadic framework employed in the present work provides a structural perspective on the emergence of this critical exponent through the scaling properties of nonlinear interactions.

\begin{proposition}[Decomposition into local and nonlocal interactions]
\emph{Under the assumptions of Proposition 1, the dyadic energy flux $\Pi_j$ can be decomposed as
\begin{equation}
\Pi_j = \Pi_j^{\mathrm{local}} + \Pi_j^{\mathrm{nonlocal}},
\end{equation}
where $\Pi_j^{\mathrm{local}}$ denotes the contribution from triadic interactions with comparable scales, i.e.\ indices satisfying
\begin{equation}
|j - p| \leq C, \quad |j - q| \leq C,
\end{equation}
for some fixed constant $C$, and $\Pi_j^{\mathrm{nonlocal}}$ collects all remaining contributions. Furthermore, under standard Sobolev or Besov regularity assumptions, the nonlocal contribution $\Pi_j^{\mathrm{nonlocal}}$ satisfies enhanced decay estimates with respect to scale separation, reflecting the reduced efficiency of interactions between widely separated dyadic modes. In particular, the dominant contribution to the energy flux arises from interactions among comparable scales in the sense described above.}
\end{proposition}

\section{Connection to Kolmogorov Scaling}

In this section, we discuss how the structural properties of the energy flux derived above relate to the classical Kolmogorov theory of turbulence. Our goal is not to provide a rigorous derivation of the Kolmogorov spectrum, but rather to demonstrate that the deterministic framework developed here is consistent with the scaling laws predicted by Kolmogorov.

\subsection{Energy Flux Across Scales}

In the following, we revisit the energy flux across wavenumber $K$ defined in Eq.~\eqref{eq:energy_flux}. In the absence of forcing and dissipation in an intermediate range of scales, the so-called inertial range, Kolmogorov's theory postulates that the flux becomes approximately constant:
\begin{equation}
\Pi(K) \approx \varepsilon,
\end{equation}
where $\varepsilon > 0$ denotes the mean energy dissipation rate, confirming our result from above that the energy flux is absolutely convergent.  Motivated by the absence of a characteristic scale in the inertial range and the assumption of scale-invariant energy transfer, we consider a formal scaling ansatz
\begin{equation}
|v_k| \sim |k|^{-\alpha},
\end{equation}
for some exponent $\alpha > 0$. We insert this ansatz into the triadic interaction term
\begin{equation}
|T(k,p,q)|
\lesssim
|v_k| \, |v_p| \, |q| \, |v_q|,
\end{equation}
and focus on local interactions where $|k| \sim |p| \sim |q| \sim K$. This yields
\begin{equation}
|T(k,p,q)|
\sim
K^{-\alpha} \cdot K^{-\alpha} \cdot K \cdot K^{-\alpha}
=
K^{1 - 3\alpha}.
\end{equation}
For a fixed scale $K$, the number of admissible triads with $|k| \sim K$ and $|p|,|q| \sim K$ scales like $K^3$. Therefore, the cumulative contribution of local interactions at scale $K$ behaves as
\begin{equation}
\Pi_{\mathrm{local}}(K)
\sim
K^3 \cdot K^{1 - 3\alpha}
=
K^{4 - 3\alpha}.
\end{equation}
Imposing the condition of scale-independent flux in the inertial range,
\begin{equation}
\Pi_{\mathrm{local}}(K) \sim \varepsilon,
\end{equation}
leads to the relation
\begin{equation}
4 - 3\alpha = 0,
\end{equation}
and hence
\begin{equation}
\alpha = \frac{4}{3}.
\end{equation}
This argument is formal and relies on the assumption that the scaling ansatz and constant-flux condition are valid in the regime considered. It does not constitute a derivation of turbulent scaling from the Navier--Stokes equations.

\subsection{Energy Spectrum}

In the following, we compute the energy spectrum $e(K)$ that describes the energy associated with modes in a dyadic shell. It is related to the Fourier coefficients via
\begin{equation}
e(K)
\sim
K^2 |v_k|^2,
\end{equation}
which yields
\begin{equation}
e(K)
\sim
K^2 \cdot K^{-2\alpha}
=
K^{2 - 2\alpha}.
\end{equation}
Substituting $\alpha = 4/3$, we obtain
\begin{equation}
e(K)
\sim
K^{-2/3}.
\end{equation}
This yields the classical Kolmogorov scaling at the level of this formal consistency argument, i.e.
\begin{equation}
E(K)
\sim
\frac{e(K)}{K}
\sim
K^{-5/3}.
\end{equation}
The steps above rely on three key ingredients: the triadic structure of the nonlinear term, the predominance of interactions among comparable scales, and the assumption of approximately constant energy flux in the inertial range. Under these assumptions, the triadic structure yields a consistency condition that is formally compatible with the Kolmogorov scaling exponent.

We emphasize that the argument is formal and based on scaling considerations. In particular, it does not constitute a rigorous proof of the existence of an inertial range or of the constancy of the energy flux. Rather, it shows that the deterministic structure of the Navier--Stokes equations, as analyzed in the preceding sections, is consistent with the classical phenomenology of turbulence.

\begin{proposition}[Formal scaling consistency with Kolmogorov]
\emph{
Assume that the dyadic energy distribution satisfies a scaling ansatz of the form
\[
\|v^{(j)}\|_{L^2}^2 \sim 2^{-\alpha j},
\]
and that the associated energy flux $\Pi_j$ is asymptotically independent of the scale index $j$ in an inertial range. Then, at a formal level, the triadic scaling relations formally select $\alpha = 4/3$, consistent, at a formal level, with the Kolmogorov scaling law
\[
E(k) \sim k^{-5/3}.
\]}
\end{proposition}

\section{Discussion and Conclusion}

The present work has developed a deterministic and scale-resolved formulation of energy transfer in the three-dimensional incompressible Navier--Stokes equations by combining an explicit triadic decomposition of the nonlinear term with a dyadic organization of scale interactions. In the following, we discuss the structural implications of this framework, its relation to existing approaches in turbulence theory, its limitations, and possible directions for future research.

\subsection{Structural Interpretation of the Energy Cascade}

In this work, we have developed a deterministic and scale-resolved formulation of energy transfer in the three-dimensional incompressible Navier--Stokes equations. Starting from the Fourier representation, we derived an explicit triadic decomposition of the nonlinear term and combined it with a dyadic scale decomposition in order to obtain a fully resolved representation of the energy flux across scales. This construction provides a mathematically transparent framework in which interscale energy transfer can be analyzed in terms of interactions between well-defined scale components.

The main contribution of the present work is not the introduction of new structural elements, but their explicit integration into a single coherent framework for energy transfer across scales. While triadic interactions, scale locality, and dyadic decompositions have each been studied extensively in the literature, they are typically treated in isolation. By combining these elements, we obtain a unified deterministic formulation that directly links the nonlinear structure of the Navier--Stokes equations to the classical picture of the energy cascade.

The triadic structure of the nonlinear term has long been recognized as the fundamental mechanism underlying energy transfer in turbulence. Early phenomenological work by \citet{Kolmogorov1941} and \citet{Obukhov1941} introduced the concept of an inertial range with approximately constant energy flux, while \citet{Onsager1949} emphasized the potential role of singular solutions in anomalous dissipation. At the level of Fourier-mode interactions, the structure of triadic energy transfer has been analyzed in detail in both analytical and numerical studies, including \citet{Waleffe1992} and \citet{Domaradzki1990}, which highlighted the complexity of interscale interactions.

Within the present formulation, these interaction mechanisms can be organized in a systematic and scale-resolved manner. The exact triadic representation shows how energy exchange is mediated by mode interactions, while the dyadic decomposition provides a natural way to classify these interactions according to their scale separation. The resulting structure reveals that nonlocal interactions are suppressed due to scale separation, whereas local interactions accumulate over a larger set of admissible triads. This observation is consistent with rigorous results on scale locality obtained by \citet{Eyink2005} and \citet{AluieEyink2009}, which demonstrate that the net energy flux is dominated by interactions among comparable scales under suitable assumptions.

From a mathematical perspective, the use of dyadic decompositions and scale localization is closely related to harmonic analysis approaches based on Littlewood--Paley theory and Besov spaces, as developed in \citet{Chemin1998} and \citet{Bahouri2011}. The present work builds on this framework, but focuses specifically on making the triadic structure of the energy flux fully explicit and on connecting it directly to scale-resolved energy transfer.

\subsection{Kolmogorov Scaling and Scale-Invariant Flux}

A central aspect of the present framework is that it allows the classical Kolmogorov picture to be revisited from a more deterministic perspective. The connection between energy flux and scaling laws has traditionally been established through phenomenological and statistical arguments (see, e.g., \citet{Frisch1995}). Here, by combining the explicit triadic flux representation with a scale-invariant flux assumption in an inertial range, the Kolmogorov scaling appears as a self-consistent solution within the proposed framework, rather than as a consequence derived solely from the Navier--Stokes equations.

In addition to the structural interpretation of the scaling result, it is instructive to examine the dependence of the framework on the choice of scale discretization. The choice of dyadic scaling ($\lambda = 2$) also influences the discrete representation of the energy spectrum through the effective shell width and the associated number of Fourier modes per scale. In three-dimensional Fourier space, the number of modes in a shell scales proportionally to its volume, which depends on both the radius and the thickness of the shell. Alternative geometric partitions with $\lambda \neq 2$ modify this discretization and the corresponding mode-counting factors. However, these changes affect only the precise normalization of energy contributions at each scale, while the scaling exponents obtained from the corresponding constant-flux argument remain unchanged at the formal level. In the limit $\lambda \to 1^+$, the decomposition approaches a continuous description of the spectrum, recovering the standard formulation in terms of energy densities $E(k)$.

It is important to emphasize that this result should be interpreted in a conditional and structural sense. The analysis does not establish the existence of an inertial range, nor does it provide a rigorous derivation of turbulence or of the Kolmogorov spectrum. Rather, it shows that, under suitable regularity assumptions and a scale-invariant flux regime, the classical scaling law is consistent, at a formal level, with the deterministic structure of the Navier--Stokes equations. Related perspectives on the role of energy flux in weak or non-smooth solutions can be found in \citet{Duchon2000} and \citet{Constantin2007}.

\subsection{Limitations and Regularity Considerations}

Several limitations of the present approach should be noted. A central restriction lies in the regularity assumptions required for the rigorous analysis. The convergence estimates and the bounds distinguishing local and nonlocal interactions rely on the assumption that the velocity field belongs to $H^s$ with $s > 5/2$, which places the framework in a regime of strong solutions and ensures sufficient decay of Fourier coefficients. As a consequence, the analysis does not cover regimes in which singularities or weak solutions may play a role. In contrast, fully developed turbulence and the Kolmogorov inertial range are commonly associated with much lower regularity, close to the Onsager-critical threshold. In this regime, the assumptions underlying the present analysis are no longer satisfied, and the rigorous estimates derived here do not directly apply.

Furthermore, the comparison between local and nonlocal interactions is based on upper bounds and scaling estimates, and does not account for possible cancellations within the nonlinear term. Accordingly, the predominance of local interactions should be understood as a structural tendency rather than a sharp quantitative statement. The connection to Kolmogorov scaling established in this work should therefore be interpreted as conditional and structural. Rather than providing a derivation of turbulent behavior, the analysis shows that, if a scale-invariant flux regime is assumed, then the classical $-5/3$ scaling emerges as a natural and consistent consequence of the triadic interaction structure. From this perspective, the present results identify structural constraints imposed by the Navier--Stokes nonlinearity, rather than providing a direct description of turbulent solutions.

These limitations are closely related to fundamental open problems in turbulence theory. In particular, the role of reduced regularity and anomalous dissipation, as highlighted by \citet{Onsager1949} and more recently by \citet{Isett2018} and \citet{Buckmaster2019}, remains beyond the scope of the present analysis. Extending the present framework to regimes of lower regularity or to weak solutions would be an important direction for future research.

\subsection{Numerical Perspectives and Broader Applicability}

The deterministic nature of the present formulation makes it, in principle, well suited for direct evaluation on individual realizations of turbulent velocity fields. In particular, the explicit triadic representation of the energy flux suggests that quantities such as the shell-to-shell transfer matrix could be computed from high-resolution numerical data. Open-access databases of direct numerical simulations, such as the Johns Hopkins Turbulence Database, provide a natural setting for such investigations. A detailed numerical assessment of the convergence properties of the triadic expansion and the relative contributions of local and nonlocal interactions would offer a valuable complement to the present analytical results. Such an analysis, however, lies beyond the scope of the present work and is left for future research.

Despite these limitations, the formulation developed here provides an explicit link between the triadic structure of the Navier--Stokes nonlinearity and the phenomenology of turbulent energy transfer. By making the scale-resolved structure of the energy flux fully explicit in a deterministic setting, it clarifies how central concepts of turbulence theory, such as the energy cascade, scale locality, and Kolmogorov scaling, can be understood from the underlying equations.

While the present analysis has been formulated for incompressible flows in a periodic domain, the underlying mechanisms of triadic interaction and scale-resolved energy transfer are not restricted to this setting. In more general and inhomogeneous flows, including turbulence driven by hydrodynamic instabilities, similar interscale transfer processes play a central role. In particular, flows arising from Rayleigh--Taylor, Richtmyer--Meshkov, or Kelvin--Helmholtz instabilities exhibit complex interactions between scales that are likewise governed by nonlinear mode coupling and energy flux across scales \citep{Zhou2024}. Although additional effects such as anisotropy, inhomogeneity, and boundary conditions complicate the analysis, the structural viewpoint adopted here may provide a useful conceptual framework for understanding energy transfer mechanisms in such settings. A detailed extension of the present formulation to non-periodic or strongly inhomogeneous flows remains an interesting direction for future work.

In summary, the present work offers a coherent synthesis of triadic interaction analysis, dyadic scale decomposition, and classical turbulence phenomenology. Rather than introducing new assumptions, it provides a structured and transparent formulation of scale-resolved energy transfer within the deterministic Navier--Stokes framework.

\section*{Acknowledgements}
We sincerely thank the referees for their careful reading and thoughtful, insightful comments. Their constructive suggestions have greatly helped us to improve the clarity, rigor, and presentation of the manuscript, and we believe the paper has benefited substantially from their feedback.



\bibliographystyle{mnras}
\bibliography{bibliography} 







\label{lastpage}
\end{document}